\documentclass[twocolumn,pra,showpacs,amsmath,amssymb]{revtex4-1}
\usepackage{physics}
\usepackage{graphicx}
\usepackage{mwe}
\usepackage{bm}
\usepackage{epstopdf}
\usepackage{epsfig}
\usepackage{dsfont}
\usepackage{amssymb}
\usepackage{amsfonts}
\usepackage[percent]{overpic}
\usepackage{subfig}
\usepackage{ hyperref }
\begin{document}

\title{Harmonic oscillator kicked by spin measurements: a Floquet-like system without classical analogous}
\author{Bento Montenegro, Nadja K. Bernardes, and Fernando Parisio}
\email[]{fernando.parisio@ufpe.br}
\affiliation{Departamento de
F\'{\i}sica, Universidade Federal de Pernambuco, Recife, Pernambuco
50670-901 Brazil}

\begin{abstract}
We present a kicked harmonic oscillator where the impulsive driving is provided by stroboscopic measurements on an ancillary degree of freedom and not by the canonical quantization of a time-dependent Hamiltonian. The ancila is dynamically entangled with the oscillator position, while the background Hamiltonian remains static. The dynamics of this system is determined in closed analytical form, allowing for the evaluation of a properly defined Loschmidt echo, ensemble averages, and phase-space portraits. As in the case of standard Floquet systems we observe regimes with crystalline and quasicrystalline structures in phase space, resonances, and evidences of chaotic behavior, however, not originating from any classically chaotic system. 
\end{abstract}
\maketitle

\section{Introduction} A wealth physical phenomena, not reachable within the realm of autonomous systems, emerge when external driving is introduced. Classical Hamiltonian chaos \cite{alfredo}, e. g., is only possible in one dimension if a time-dependent disturbance is present. Often, these external influences are time-periodic and impulsive, as is the case of the archetypal delta-kicked mechanical systems. Their quantum versions, such as tops \cite{TOP1,TOP2,TOP3,TOP4} and oscillators \cite{OH0, OH1, OH2}, among others \cite{alfredo2}, are, arguably, even more intriguing, not only under the perspective of quantum chaos \cite{QC}. These systems have been shown to display quantum phase transitions, fractal bands in the quasi-energy spectrum, and crystalline patterns in both phase space and time domain. 

Quantum mechanically, whenever a Hamiltonian satisfies the periodicity condition $\hat{H}(t)=\hat{H}(t+T)$, one has a Floquet  system \cite{floquet1,floquet2}. Usualy, it is assumed that the disturbance comes from the canonical quantization of a time-dependent Hamiltonian, often containing a comb of impulsive terms.  Quantum dynamics, however, is not exhausted by the Sch\"odinger evolution. It also comprises the abrupt changes caused by measurements, impulsive disturbances par excellence, with no classical parallel. 

The question arises, is it possible to augment the set of quantum systems that can develop a Floquet-like dynamics by using measurements as a {\it dynamical ingredient}? The answer is positive if one employs time-periodic (stroboscopic) measurements \cite{Von}. If, however, these measurements directly refer to, say, a particle position, the dynamics becomes trivial, since they would only ``reset'' the system at a random position, after each projection. If, instead, we use an auxiliary degree of freedom, correlated with the particle position, and carry out measurements on it, they work  as genuine non-unitary kicks. 

To introduce and characterize such a system, presenting evidence of quantum chaos and formation of phase-space crystals, is the purpose of this work.  We remark that this approach is distinct from those which derive time-independent {\it effective} Hamiltonians for Floquet systems \cite{effec1,effec2,effec3,effec4,effec5,effec6}, usually under a fast-driving condition. 

In Sections II and III, we describe the system to be addressed and provide a closed analytical solution for its dynamics, respectively. In section IV, we show that the system displays dynamical instability, namely,  exponential Loschmidt echo decay, suggesting quantum chaos. 
The route for this non-trivial dynamics may represent new terrain in the field of quantum dynamical systems, since it stems from the invasiveness of measurements, thus, having a non-Hamiltonian character. Section V is devoted to the derivation of several ensemble averages, while section VI addresses phase-space crystals. In this regard, to have a system bounded by a single well and, at the same time, a periodic structure in phase-space, means that one can artificially build a physical system with properties of a solid, at least in some respects. These properties can be qualitatively changed, e. g., from crystalline to vitreous, passing through quasi-crystals and back to a crystalline form, by varying an experimentally controllable parameter.  In the last section we give our final remarks. To highlight the key results and improve readability, we deferred the demonstrations of most of the technical developments to a series of appendices.
\section{System} 
\begin{figure}[h]
		\includegraphics[height=5.2cm,angle=0]{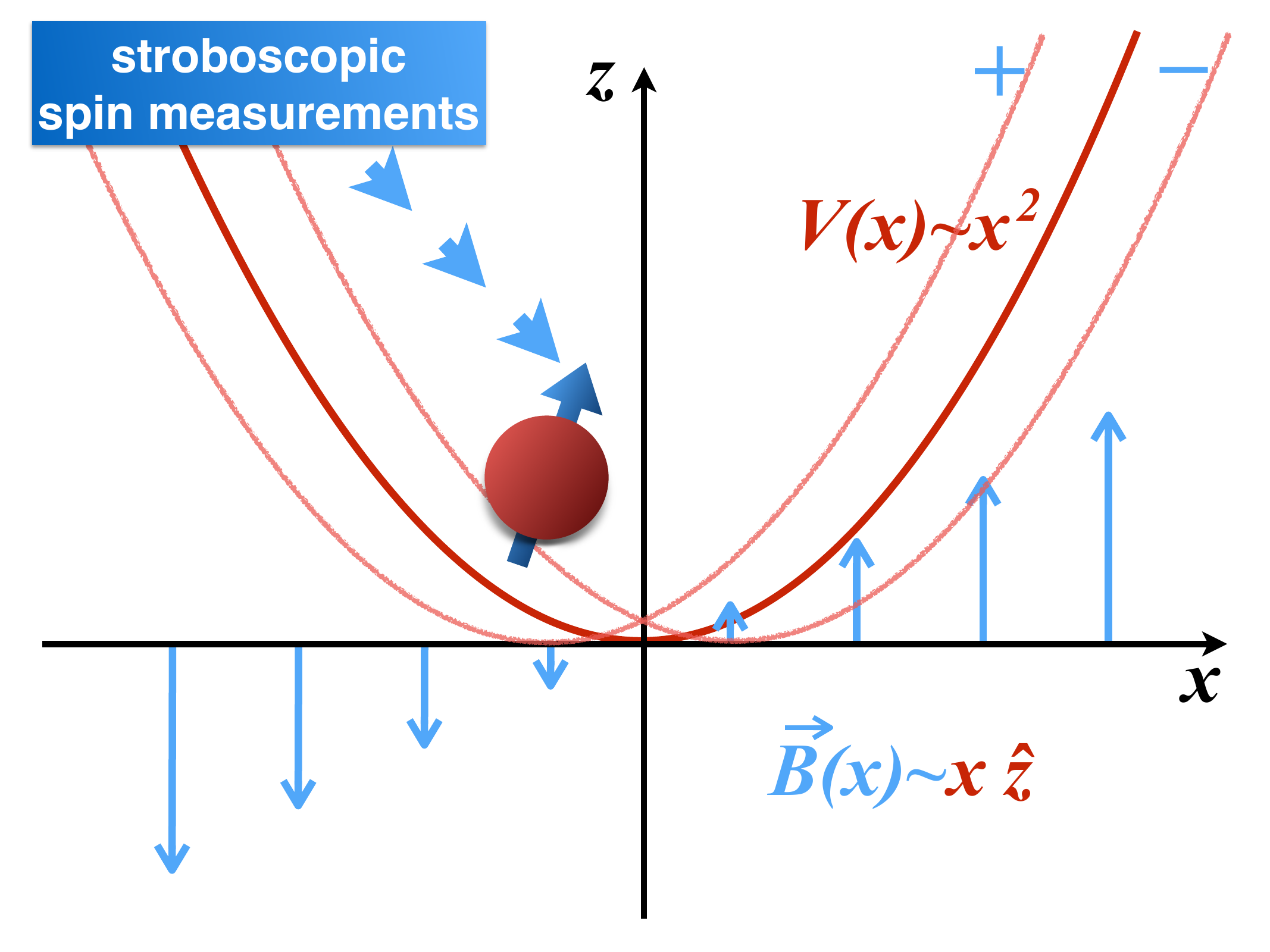}
		\caption{(color online) The magnetic field $\vec{B}(x)\sim x \hat{z}$ couples position (red) and spin (blue). $\hat{S}_x$ measurements periodically project the system into a non-stationary state.  The branch $\ket*{z}\ket*{+}$ ($\ket*{z}\ket*{-}$) ``sees'' a harmonic potential of frequency $\omega_0$ displaced to the left (right).} 
		\label{fig1}
\end{figure}

We will consider a spin-$1/2$ particle of mass $m$ and magnetic moment $\hat{\mu}=\gamma\hat{S}$, $\hat{S}$ being the spin operator. The particle motion is constrained to an effectively one dimensional region of space, the $x$ axis of our coordinate system, and bounded by a harmonic potential of frequency $\omega_0$.  If the particle is also immersed in an inhomogeneous magnetic field $\vec{B}(x, y, z)$, whose restriction to $y=0$ and $z=0$ reads 
$$\vec{B}(x, 0, 0)\equiv \vec{B}(x)=\sqrt{2}B_0 \left(\frac{x}{b}\right) \vec{e}_z,$$ 
the orbital and intrinsic degrees of freedom may become dynamically entangled, where $b = \sqrt{\hbar/m\omega_0}$ is a constant with dimension of space [there are, of course, infinitely many 3D forms of $\vec{B}(x,y,z)$, compatible with $\vec{B}(x,0,0)=B_0 (x/b) \hat{z}$ and  $\vec{\nabla} \cdot \vec{B}=0$]. This correlation will be essential in the developments of the next sections, where successive spin projective measurements will be considered (see Fig. \ref{fig1}).
The relevant Hilbert space is ${\cal H}_{X}\otimes {\cal H}_{S}$ and the Hamiltonian of the system reads
\begin{equation}
\label{OurHamiltonian}
    \hat{H} = \hat{H}_{HO} \otimes \mathds{1} - \frac{\sqrt{2}\alpha}{b} \hat{X} \otimes \hat{\sigma}_z,
\end{equation}
where  $\hat{H}_{HO} = \hat{P}^2/2m +V(\hat{X})$, $ V(\hat{X})=m \omega_0^2 \hat{X}^2/2$, is the simple harmonic oscillator (HO) Hamiltonian and $\hat{\sigma}_z$ is the  $z$-direction Pauli operator.
We mention that the fragility of quantum superpositions for an equivalent system under {\it Hamiltonian} disturbances has been addressed in reference \cite{LE-QC1b}.
 The parameter $\alpha = \gamma B_0 \hbar/2$  gives the entangling strength of position and spin. We will be interested in the phase-space dynamics given that the initial state is maximally localized, that is, a canonical (or quasi-classical) coherent state ($\ket*{z_0}$), initially uncorrelated to the spin. We thus, consider the initial state to be $\ket*{\psi(0)}=\ket*{z_0}\ket*{\chi}$, where $\ket*{\chi}$ is a convenient spin state, which must not be an eigenket of $\hat{\sigma}_z$, otherwise no spin-orbit correlation develops. In order to maximize the interaction we choose $\ket*{\chi}$ as an eigenket of $\hat{\sigma}_x$, the Pauli operator in the $x$ direction, $\hat{\sigma}_x \ket*{s}_x= s \ket*{s}_x=s(\ket*{+}+s\ket*{-})/\sqrt{2}$, $s=\pm1$ ($\ket*{\pm}$ being eigenkets of $\hat{\sigma}_z$).  

The action of $\hat{H}$ on the initial state gives $\hat{H} \ket*{\psi} =\hat{H}\ket*{z_0} \ket*{s}_x=( \hat{H}_+\ket*{z_0}\ket*{+}+s \hat{H}_-\ket*{z_0}\ket*{-})/\sqrt{2}$.
The Hamiltonians $\hat{H}_{\pm}$ act only on the orbital degree of freedom and read $$\hat{H}_\pm = \hbar \omega_0 \Big{(} \hat{a}^\dag_\pm \hat{a}_\pm + \frac{1}{2} \Big{)} - \hbar \omega_0 v^2,$$ where $v=\alpha/\hbar\omega_0$ and we have defined displaced ladder operators
$\hat{a}_\pm = \hat{a} \mp v$, with $\hat{a}$ being the canonical annihilation operator (analogously for $\hat{a}_\pm^{\dagger}$). It is clear that the initial coherent state will be split by the ``kick'', since the term $\ket*{z_0}\ket*{+}$ ($\ket*{z_0}\ket*{-}$) senses a harmonic potential, with the same frequency $\omega_0$, displaced to the left (right), see Fig. \ref{fig1}. Because the shape of the potential remains the same, each branch will evolve as a coherent state: $$\hat{U}_{\pm}^{T}\ket*{z_0}  = e^{-i \omega_0 (\frac{1}{2} - v^2) T} \ket*{(z_0 \mp v) e^{-i\omega_0T} \pm v},$$
where $ \hat{U}_{\pm}^{T}= \exp\{ i \hat{H}_{\pm}T/\hbar\}$. So, importantly, the initial coherent state evolves into a superposition of {\it coherent states}. For a fixed value of $T$, we define the ``$\pm$'' maps
\begin{equation}
\nonumber
\mathcal{Z}_{\pm}: \mathbb{C} \rightarrow \mathbb{C}\;| \;\mathcal{Z}_{\pm} (z_0)\equiv (z_0 \mp v) e^{-i\omega_0 T} \pm v.
\end{equation}
Therefore, the unitary evolution of the $\pm$ branches of the state vector, after a time $T$, is given by:
$ \frac{1}{2} [ (\ket*{\mathcal{Z} _{ +}} + s_0 \ket*{\mathcal{Z} _{-}}) \ket*{+}_x + (\ket*{\mathcal{Z} _{+}} - s_0 \ket*{\mathcal{Z} _{-}}) \ket*{-}_x)]$,
where, for future convenience, we wrote the spin part in terms of eigenstates of $\hat{\sigma}_x$ and also skipped the global phase factor.
The general form of $N$ arbitrary compositions of the maps $+$ and $-$ can be derived in closed form and reads:
\begin{equation}\label{GeneralMap}
\mathcal{Z}_{I_N} \hspace{-0.08cm}= \hspace{-0.1cm}\left(\hspace{-0.08cm}z_0 + 2 i v e^{-\frac{i \omega_0 T}{2}}\hspace{-0.05cm} \sin \left( \frac{\omega_0 T}{2} \right)\hspace{-0.1cm} \sum_{j=1}^{N} i_j e^{i  \omega_0 T\, j} \hspace{-0.08cm} \right)\hspace{-0.1cm} e^{- i N \omega_0 T}.
\end{equation}
In this notation, $I_N = \{i_N, i_{N-1}, ..., i_2, i_1\}$ denotes a specific arrangement of $N$ elements $i_k = \pm 1$. The map $\mathcal{Z}_{i_1}$ is applied first, followed by the maps $\mathcal{Z}_{i_2}$, $\mathcal{Z}_{i_3}$, ..., $\mathcal{Z}_{i_N}$. Equation (\ref{GeneralMap}) is demonstrated in appendix \ref{Zmap} and its use will become clear soon. 

\section{Quantum trajectories} 
We are now in a position to introduce the stroboscopic spin projections, which we assume to be in the $x$ direction. Since the restriction of the magnetic field to the $x$ axis points in the $z$ direction, the system undergoes an entangling evolution between successive measurements of $\hat{S}_x=\hbar \hat{\sigma}_x/2$, separated by a time $T$, corresponding to a stroboscopic angular frequency of $\omega=2\pi/T$. The other relevant time scales are the HO period $T_0=2\pi/\omega_0$ and the Larmor period $T_L=2\pi/\gamma B_0$ of the spin. The frequency ratio $R \equiv \omega_0/\omega$ is shown to be particularly relevant. For each $\hat{S}_x$ measurement we have two possible results, so, after $N$ sequential measurements, there are $2^N$ possible sets of outcomes. We will refer to each of these sets and to the states the system is projected in the process, a quantum trajectory \cite{QT1,QT2,QT3,QT4,QT5}. We denote the set of all quantum trajectories after $N$ measurements as $\gamma^N$, and a specific trajectory as $\gamma^N_k$, $k=1,2, 3, \cdots, 2^N$. Each trajectory, in turn, is fully specified by the records of all $\hat{\sigma}_x$ measurements, denoted by $\{s_j^k\}=\{ s_1^k, s_2^k, \cdots, s_N^k\}$. In appendix  \ref{sec:time-evolution} we calculate the final orbital state of the system for an arbitrary quantum trajectory  $\gamma^N_k$:
\begin{equation}\label{PsiN}
\ket*{\psi_N^k} = \frac{\sum_{I_N} c_{I_N}^k \ket*{\mathcal{Z}_{I_N}}}{\norm{\sum_{I_N} c_{I_N}^k \ket*{\mathcal{Z}_{I_N}}}}  ,
\end{equation}
with $c_{I_N}^k \equiv \prod_{j=1}^{N} (s_m^k s_{m-1}^k)^{\delta_{i_m,-1}} = \pm 1$, where
$\delta_{i,j}$ is the Kronecker delta symbol. We note that the discrete time-evolution of each state $ \ket*{\psi_N^k}$ is defined on
a topological structure akin to a 3-coordinated Bethe lattice \cite{bethe,bethe2} in phase-space, each node corresponding to
one particular realization of (\ref{GeneralMap}).
The fact that the phase ($\pm1$) attributed to each node $\ket*{\mathcal{Z}_{I_N}}$ depends on the outcome 
of spin measurements resembles the original implementation of a quantum random walk \cite{RW} (see also \cite{RW1,RW2}),
where the displacement depends on the same kind of projection.
The probability for each trajectory to occur is given by $p_N[\gamma_k^N] = p_1(s_1^k|s_0) p_2(s_2^k|s_1^k,s_0)...p_N(s_N^k|s_{N-1}^k,...,s_0)$, which reads:
\begin{equation}\label{probtraj}
p_N[\gamma_k^N] = 4^{-N}\norm{ \sum_{I_N} c_{I_N}^k \ket*{\mathcal{Z}_{I_N}} }^2.
\end{equation}
Equations (\ref{PsiN}) and (\ref{probtraj}) are demonstrated in appendix \ref{sec:time-evolution}.

\begin{figure}
 \setbox1=\hbox{\includegraphics[height=3.2cm]{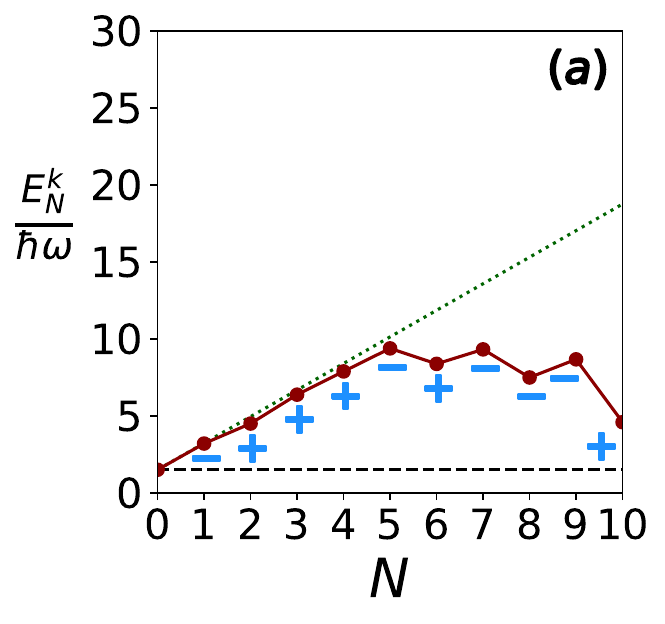}}
  \includegraphics[height=4.2cm]{fig2a.pdf}\llap{\makebox[\wd1][l]{\raisebox{2.28cm}{\includegraphics[height=1.7cm]{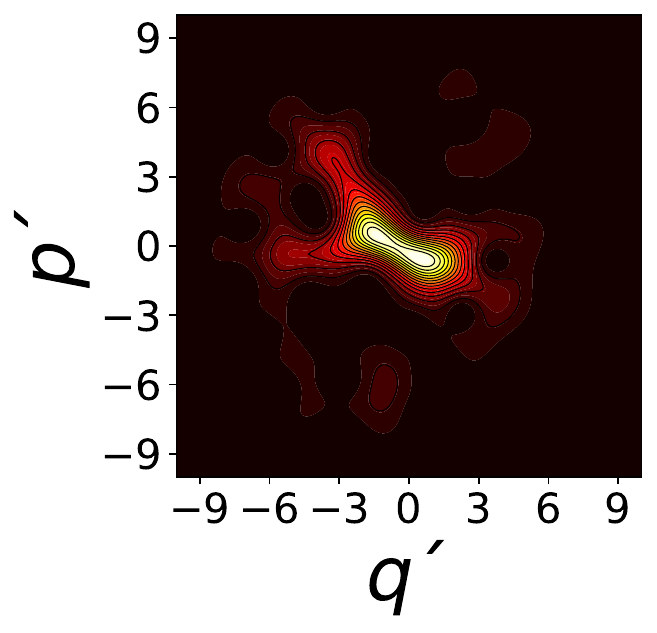}}}}
  \setbox1=\hbox{\includegraphics[height=3.6cm]{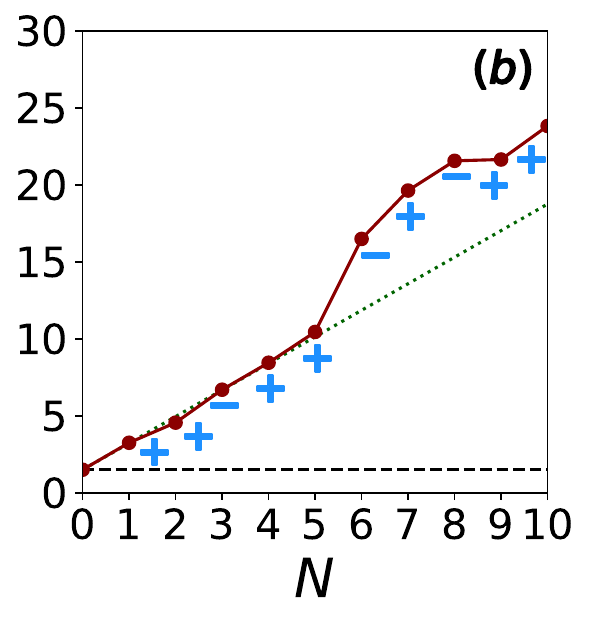}}
  \includegraphics[height=4.2cm]{fig2b.pdf}\llap{\makebox[\wd1][l]{\raisebox{2.28cm}{\includegraphics[height=1.65cm]{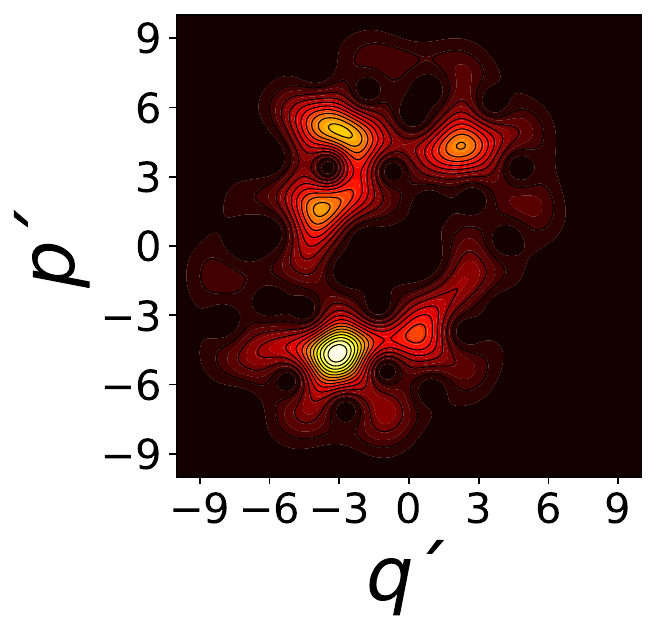}}}}
  \caption{(color online) Energy expectation values for two quantum trajectories determined by the outcomes of $\hat{S}_x$ measurements (blue signs). The insets show the associated Husimi functions for $N=6$. Darker (lighter) regions correspond to lower (higher) probabilities. We used $v=2.0$, $z_0=(1+i)/\sqrt{2}$, and $R=0.106$.}
  \label{fig2}
\end{figure}

With these results we can calculate the energy expectation value ($E_N^k= \bra*{\psi_N^k} \hat{H}  \ket*{\psi_N^k}$) and phase-space portraits, e. g., by means 
of the Husimi function: $h_N^k(q,p)=|\bra*{z}\ket*{\psi_N^k}|^2$, where $\ket*{z}$ is a coherent state with 
$z=(q/b+ibp/\hbar)/\sqrt{2} \equiv q'+ip'$ ($q'$ and $p'$ being dimensionless position and momentum variables). In Fig. \ref{fig2} the values of $E_N^k/\hbar\omega_0$ are depicted as a function of the discrete-time variable $N$, for two illustrative quantum trajectories, characterized by the displayed sequence of signs. Notice that the results fluctuate in an erratic manner, particularly for $N \ge 6$. In the left panel we observe that the energy can drop between successive measurements, for a sufficiently small $R$ ($R=0.106$), however, this is not a typical behavior, for, the average over all trajectories leads to a steady, linear energy increase as $N$ grows (dotted straight lines), see Eq. (\ref{Hclosed}) below. The insets show the Husimi functions for $N=6$. Note how different the phase space portraits may be for different quantum trajectories.
\section{Dynamical instability}
The irregular behavior of individual trajectories suggests dynamical sensitivity and the possibility of quantum chaos. The most direct quantity we may use to investigate this hypothesis is the Loschmidt echo \cite{LE,LE1}.
This quantity has been repeatedly employed in the study of the dynamical stability of quantum systems, including delta-kicked ones \cite{LE-QC,LE-QC1,LE-QC1b,LE-QC2}. The Loschmidt echo has also been shown to be tightly connected with other quantum chaos diagnosis tools, as is the case of out-of-time-order correlator, see \cite{LE-QC3} and references therein. 

We must, however, make the echo meaningful for the peculiar system we are addressing. First, we should exclude from the analysis the randomness introduced by the quantum measurements. Therefore, we will compare equal initial states evolved through slightly different Hamiltonians {\it and} for which all spin measurements turned out to yield the same result. Thus, we define the echo $L^{k}_N=|\langle \psi_N^k|{\psi'}_N^k\rangle|^2$ for a fixed set of spin outputs $\gamma_N^k$, where $|{\psi'}_N^k\rangle$ is the state evolved in a harmonic potential with natural frequency $\omega_0+\delta \omega_0$. In addition, in order to compare identical initial coherent states (same $q_0$, $p_0$, and $b$), and still be able to use (\ref{GeneralMap}), this frequency change must be accompanied by a change in the mass, such that $m\omega_0=(m-\delta m)(\omega_0+\delta \omega_0)$, which leaves the parameter $b=\sqrt{\hbar/m\omega_0}$ unchanged in both realizations. These two slight modifications can be seen as a small disturbance in the original Hamiltonian. 
\begin{figure}
 \setbox1=\hbox{\includegraphics[height=2.25cm]{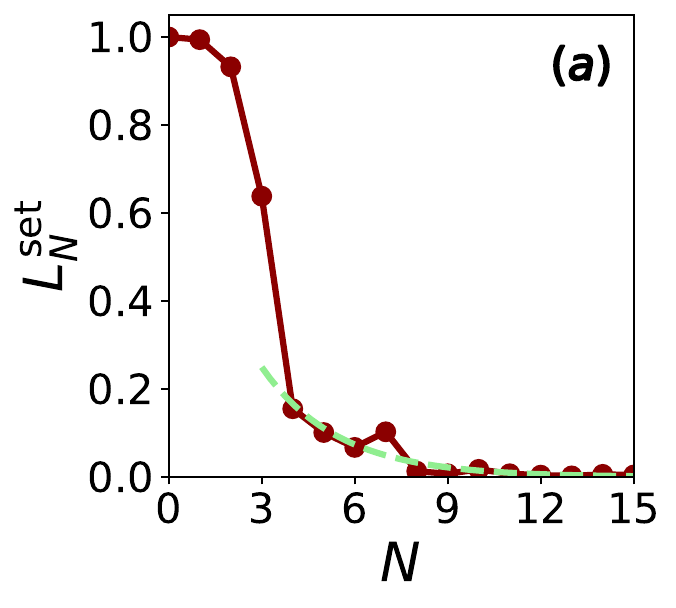}}
  \includegraphics[height=4cm]{fig3a.pdf}\llap{\makebox[\wd1][l]{\raisebox{1.5cm}{\includegraphics[height=1.85cm]{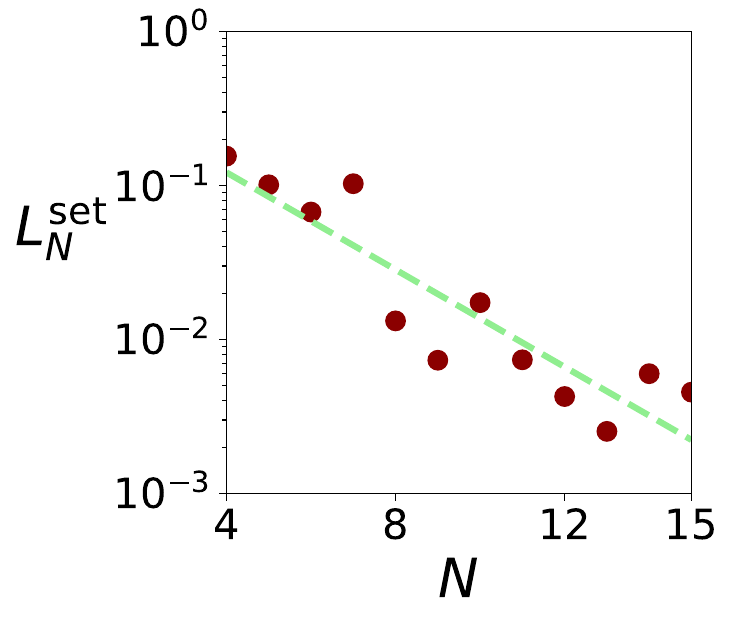}}}}
  \setbox1=\hbox{\includegraphics[height=2.45cm]{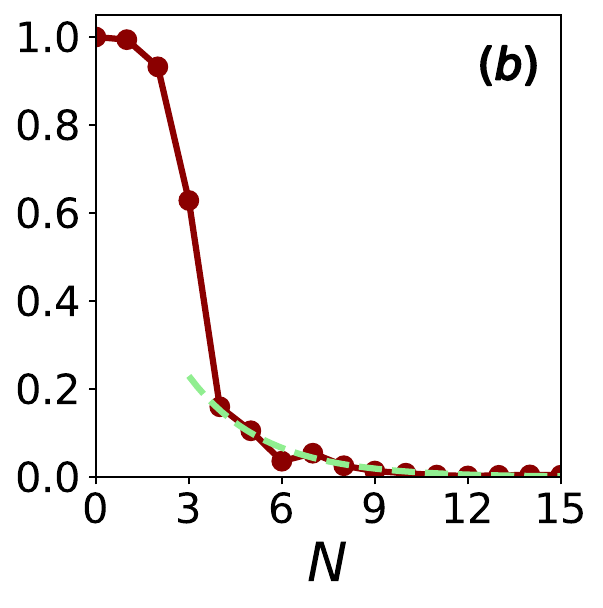}}
  \includegraphics[height=4cm]{fig3b.pdf}\llap{\makebox[\wd1][l]{\raisebox{1.5cm}{\includegraphics[height=1.85cm]{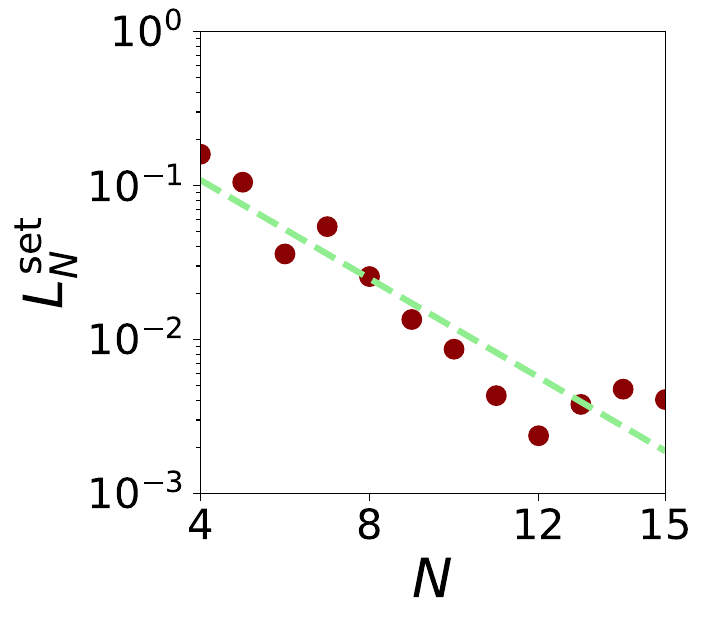}}}}
  \caption{(color online) The averaged Loschmidt echo decay for two independent sets of 15 quantum trajectories, (a) and (b).  We used the same parameters as in Fig. \ref{fig2} except for $R=1/5$. For the small perturbation in the Hamiltonian we set $\delta R=5/1000$. The inset shows log-scale plots for $N>3$. The exponential fitting is given by the green dashed lines.}
  \label{fig3}
\end{figure}

In order to investigate the functional dependence of the echo decay, in a statistically significant way, we picked sets of 15 quantum trajectories and calculated the average Loschmidt echo up to the discrete time of $N=15$. Each trajectory is obtained by drawing 15 random signs ($+$ or $-$) and the corresponding probability of occurrence is calculated from Eq. (\ref{probtraj}). 
The echo of each single trajectoriy was calculated with a disturbed Hamiltonian for the dynamics of $|{\psi'}_N^k\rangle$ with $R=0.2$ and $\delta R=0.005$, the other parameters being the same as in Fig. \ref{fig2}. The variation $\delta R$ is taken with a fixed stroboscopic frequency ($\delta R=\delta \omega_0/\omega$).
The echoes associated with the 15 trajectories were averaged:
$$L^{\rm set}_N=\frac{1}{P}\sum_{k=1}^{15}p_N[\gamma_k^N]\,L^{k}_N,$$
with $P=\sum_{k=1}^{15}p_N[\gamma_k^N]$. The corresponding results are shown in figure \ref{fig3} for two distinct sets of 15 random trajectories. We found that, after a fast transient quadratic decay, the Loschmidt echo presents an exponential drop 
 $$L^{\rm set}_N\sim e^{-\Gamma N},$$ for $N>3$, as it can be seen from the log plots in the insets of figures \ref{fig3}(a) and \ref{fig3}(b), the values of $\Gamma$ being $0.364$ and $0.368$, respectively. While the initial quadratic decay is a universal property, derived by Peres \cite{LE}, the subsequent exponential decay constitutes strong evidence of quantum chaos \cite{LE3}. However, more extensive numerical tests are desirable. Another feature worth of further investigation is the transition between the two mentioned regimes, corresponding to $N=3$, for the parameters we used.  
 
If all trajectories were equally likely a probability of $2^{-N}=2^{-15}$ would be attached to each $\gamma_k^{15}$. Therefore, in such a uniform case, the probability of ocurrence of any set of 15 trajectories would be $15/2^{15}\approx 0.00046$. The sets of trajectories we considered have probabilities of the compatible magnitude. Finally, we remark that the same set of signs usually give rise to distinct trajectory probabilities when the non-disturbed and disturbed Hamiltonian are used, corresponding to $p_N[\gamma_k^N]$ and $p'_N[\gamma_k^N]$. In the sets we considered the modular difference $|p_N[\gamma_k^N]-p'_N[\gamma_k^N]|$ is one order of magnitude smaller than $p_N[\gamma_k^N]$, on  average. We have investigated other two sets of trajectories with nearly indistinguishable results.

\section{Ensemble averages} 
Since we have determined the time evolution of every trajectory, we may gather (\ref{PsiN}) and (\ref{probtraj}) to compose the full ensamble state $\rho_N=\sum_k p_N[\gamma_k^N]  \ket*{\psi_N^k}  \bra*{\psi_N^k}$:
\begin{equation}\label{ReducedStateProtocol}
    \hat{\rho}_N = 2^{-N} \sum_{I_N}  \ket*{\mathcal{Z}_{I_N}} \bra*{\mathcal{Z}_{I_N}}.
\end{equation}
Note that none of the final states for the quantum trajectories is given by $\ket*{\mathcal{Z}_{I_N}}$, but the final state of the ensamble is an equiprobable mixture of them, with each 
$\ket*{\mathcal{Z}_{I_N}}$ representing an effective microcanonical state. We remark that Eqs. (\ref{PsiN}) and (\ref{ReducedStateProtocol}) are the analogous of the Floquet theorem for systems with periodically driven Hamiltonians, that is, the explicit expression of the time-evolved states at the stroboscopic times $NT$.
We are, thus, able to explicitly calculate the ensamble averages of energy, position, momentum, and their variances.

In appendix \ref{ensembleAP} we show that the ensemble average of the energy is given by:
\begin{equation}\label{Hclosed}
    \langle \hat{H} \rangle_N = \hbar \omega_0\left( |z_0|^2 + \frac{1}{2}  + 4v^2 \sin^2\left( \frac{\omega_0 T}{2} \right) N\right).
\end{equation}
Thus, although the system energy may drop in a particular realization,  the kicks embodied by the spin projections raise the energy by an amount of $4\hbar\omega_0 v^2 \sin^2(\omega_0T/2)$, on average (dotted lines in Fig. \ref{fig2}). This is a common regime in usual Floquet systems \cite{OH1}. For a total time $\Delta t=N T$ we write the average delivered power as
\begin{equation}
\bar{\cal P}=\frac{  \langle \hat{H} \rangle_N-  \langle \hat{H} \rangle_0}{\Delta t}= \frac{2\pi\hbar}{ R T_L^2}\sin^2\left( \pi R\right),
\end{equation}
which, as a function of the ratio $R= \omega_0/\omega$, has an absolute maximum, a resonance, for $\tan(\pi R)=2\pi R$. Numerically solving this equation, we get the stroboscopic period that maximizes power, in terms of the HO period:
$$ T_{\rm res}\approx 0.3710\, T_0 \Rightarrow \bar{\cal P}_{\rm max}\approx 16.94\, \hbar/ T_L^2.$$
So, the variable $R$ determines the resonance, while the Larmor period gives its intensity. If we take $T \ll 1$ (so that $\langle \hat{H} \rangle_N$ can be approximated by a continuous function of time), $N = t/T$, the mean power becomes 
$$ \bar{\cal P}_{\rm cont}=\frac{\partial \langle \hat{H} \rangle}{\partial t} = \frac{4\hbar \omega_0 v^2}{T} \sin^2{\left( \frac{\omega_0 T}{2} \right)} \sim T.$$
Note that in the Zeno limit $T \rightarrow 0$, we get $ \bar{\cal P}_{\rm cont} \rightarrow 0$, as it should be.
Curiously, the ensemble dynamics of the mean values of position and momentum is insensitive to both the spin-position correlation and stroboscopic measurements  (see appendix \ref{ensembleAP}), being given by: $\langle \hat{X} \rangle_N/\sqrt{2}b = [\Re{z_0} \cos{(N \omega_0 T)} + \Im{z_0} \sin{(N \omega_0 T)}$ and $b\langle \hat{P} \rangle_N/\sqrt{2}\hbar = - \Re{z_0} \sin{(N \omega_0 T)} + \Im{z_0} \cos{(N \omega_0 T)} $, which are identical to the expectation values for the simple HO. This is not incompatible with our conclusion that the system mean energy steadily increases. The energy imparted to the system is embodied by the variances $\Delta X$ and $\Delta P$, which drastically depart from those of the simple HO, being given by:
\begin{equation}\label{eq:x2_mean}
\frac{\Delta X_N^2}{ 2 b^2}=  4 v^2 \sin^2 \left(\frac{\omega_0 T}{2}\right) \sum_{\ell=1}^{2N - 1} \sin[2]( \frac{\ell \omega_0 T}{2} ) + \frac{1}{4},
\end{equation}
\begin{equation}\label{eq:p2_mean}
 \frac{b^2\Delta P_N^2}{2 \hbar^2}= 4 v^2 \sin^2 \left(\frac{\omega_0 T}{2}\right) \sum_{\ell=1}^{2N - 1} \cos[2]( \frac{\ell \omega_0 T}{2} ) + \frac{1}{4},
\end{equation}
where $N \ge 1$ and the sums are over $\ell =1,3,5,\dots$. Both variances increase with the discrete time variable $N$. These expressions are related to the average energy, Eq. (\ref{Hclosed}), through:
\begin{equation}\label{eq:xp2_mean}
\frac{b^2\Delta P_N^2}{2 \hbar^2}+\frac{\Delta X_N^2}{ 2 b^2}= \frac{ \langle \hat{H} \rangle_N}{\hbar\omega_0}-|z_0|^2.
\end{equation}
\section{Crystalline structures in phase-space} 
It is known that regular patterns in phase space may appear, also for one-dimensional systems, via {\it Hamiltonian} driving, even for particles trapped in a single potential well. These structures have been dubbed phase-space crystals and recently attracted a great deal of attention \cite{cryst, cryst2, cryst3}. We show that the presented system has the ability to display such phase-space crystals through a distinct mechanism, thus, providing a new platform to engineer systems in, e. g., many body physics \cite{cryst3}.

The Husimi function for the ensemble, $h(p,q)=\langle z|\hat{\rho}|z\rangle$, with $\rho$ given by Eq. (\ref{ReducedStateProtocol}), is a combination of $2^N$ Gaussian functions:
\begin{equation}
 h(p,q) = \frac{1}{(2 \pi \hbar)^22^N} \sum_{I_N} e^{- \frac{1}{2b^2} \qty( q - q_{I_N} )^2  - \frac{b^2}{2 \hbar^2} \qty ( p - p_{I_N} )^2 },
\end{equation}
where
$q_{I_N} = \expval{X}_N + v \sum_{j=1}^N i_j \mathfrak{q}_j$ and $p_{I_N} = \expval{P}_N + v  \sum_{j=1}^N i_j \mathfrak{p}_j$, with $\sqrt{2}\mathfrak{q}_j/b= \cos  ((N - j) \omega_0 T) - \cos ((N - j + 1) \omega_0 T)$, $b\mathfrak{p}_j/\sqrt{2}\hbar = \sin  ((N - j) \omega_0 T) - \sin ((N - j + 1) \omega_0 T)$. 
\begin{figure}[h]
	\hspace{0.08cm}
		\includegraphics[height=2.5cm,angle=0]{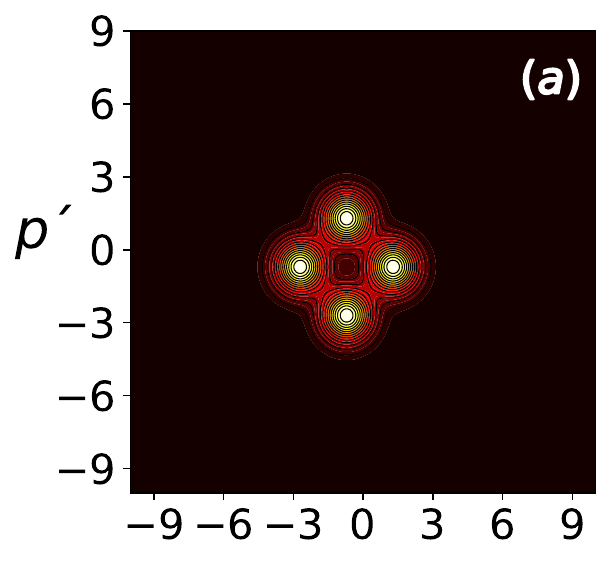}
		\hspace{0.06cm}
		\includegraphics[height=2.5cm,angle=0]{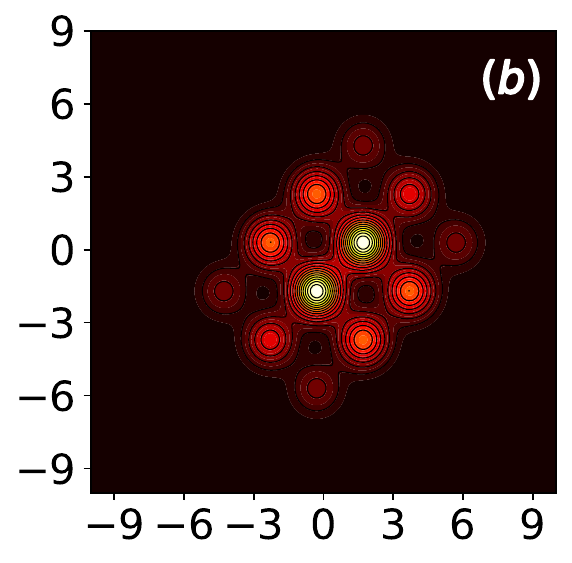}
		\hspace{0.06cm}
		\includegraphics[height=2.5cm,angle=0]{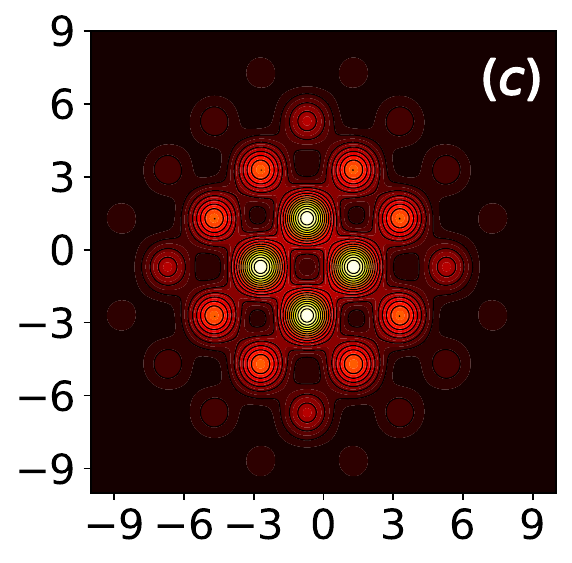}
		\includegraphics[height=2.46cm,angle=0]{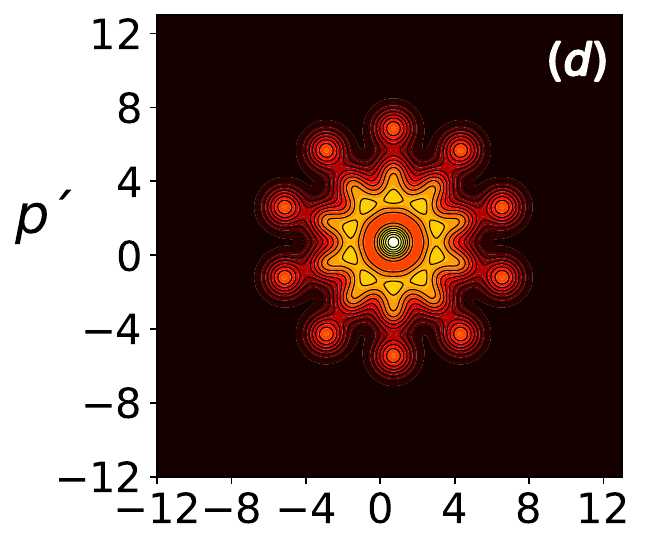}
		\includegraphics[height=2.46cm,angle=0]{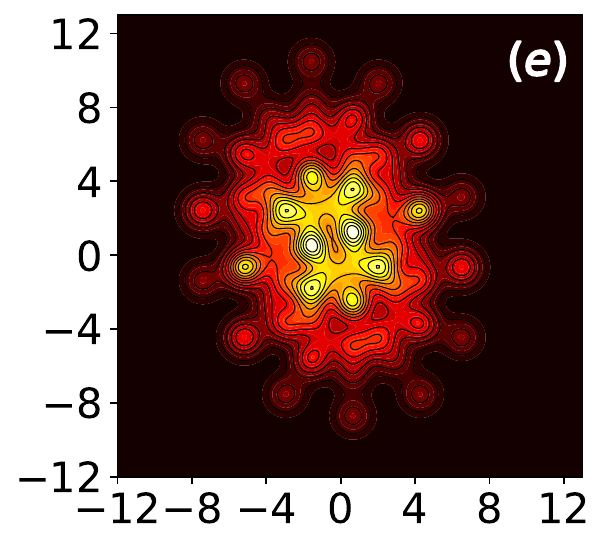}
		\includegraphics[height=2.46cm,angle=0]{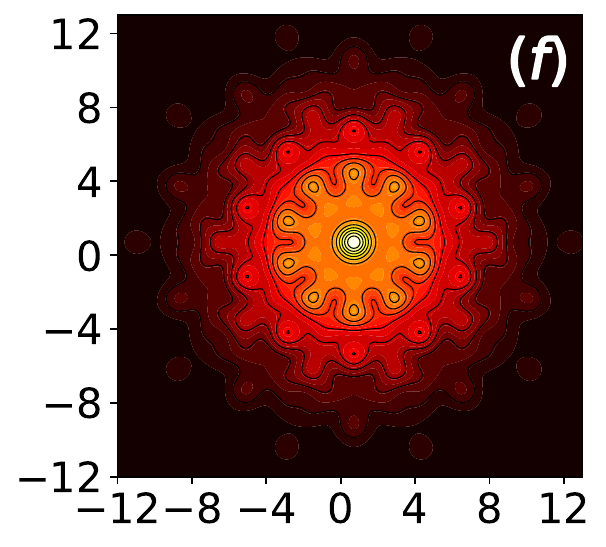}
		\includegraphics[height=2.7cm,angle=0]{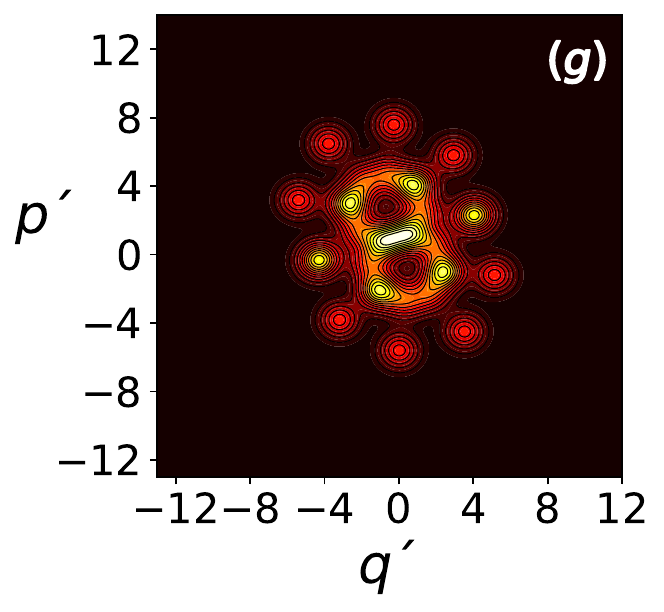}
		\includegraphics[height=2.7cm,angle=0]{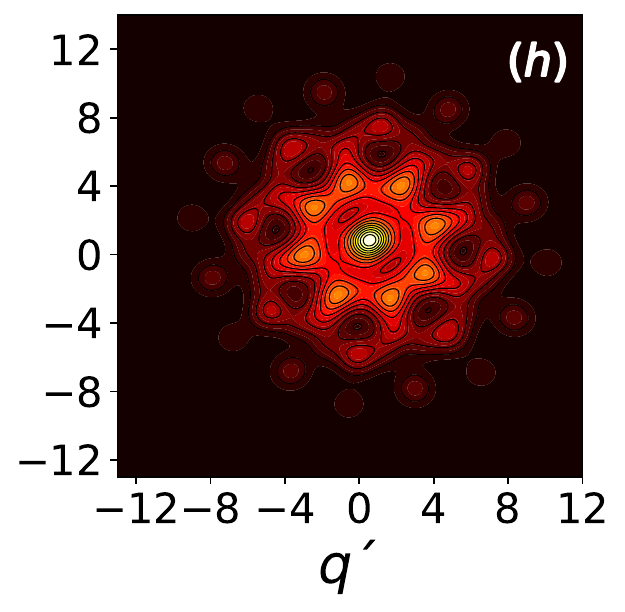}
		\includegraphics[height=2.7cm,angle=0]{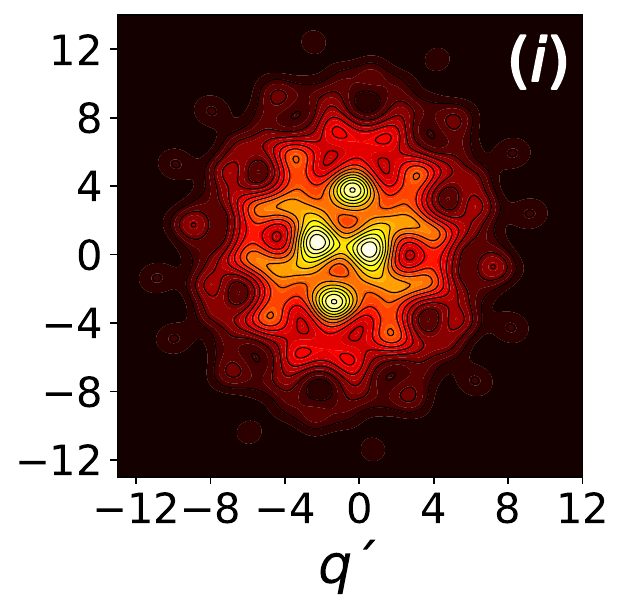}
		\caption{(color online) Husimi functions for $N=2 (a), 5 (b), 10 (c)$with $R=1/4$; $N=5(d), 7 (e), 10 (f)$ with $R=2/5$, and $N=5 (g), 8 (h),10(i)$ with the resonant frequency ratio $R\approx 0.3710$. Lighter (Darker) collors denote higher (lower) probabilities.} 
		\label{fig4}
\end{figure}
In the upper row of Fig. \ref{fig4} we show phase-space portraits for the rational frequency ratio, $R=1/4$. A crystalline, square lattice emerges and develops as the number of spin measurements increase [see the Husimi plots in Fig. 4 of  \cite{cryst3}].  
For $R=1/6$ (not shown), again a regular structure appears, this time with triangles clustering to form regular hexagons. 
For $R=2/5$, a 10-fold radially symmetric structure shows up periodically in time, see de middle row of Fig. \ref{fig4} [See Fig. 1(b) of \cite{cryst} and Fig. 4 of \cite{cryst2}]. 

For irrational values of $R$, we observe more complex structures. Consider, for instance, the transcendental resonance condition $R\approx 0.3710$. In this case, no phase-space crystal is formed, although quasi-crystalline structures appear, see the lower row of Fig. \ref{fig4}, where we observe pentagonal structures. Since these polygons are unable to cover the plane ($\mathds{R}^2$), geometric frustration takes place.

We remark that, given the nature of the ``kicks'', any of these patterns can, in principle, be frozen on demand. As soon as the desired crystal is formed one only has to either switch to a continuous-measurement regime, leading to a quasi-static Zeno dynamics, or to simply halt the stroboscopic measurements. 

In addition, these phase-space crystals do not require fine tuning, in the sense that they occur for any choice of the initial coherent state $\ket{z_0}$. This can be noticed through the maps in Eq. (\ref{GeneralMap}), for which it is clear that $z_0$ only adds an overall translation in phase space. Indeed, one can write the map as a phase-space translation ($ \mathcal{T}$) followed by an overall phase-space rotation ($\mathcal{R}$) applied to  
$$\mathcal{W}_{I_N}=2 i v e^{-\frac{i \omega_0 T}{2}} \sin \left( \frac{\omega_0 T}{2} \right) \sum_{j=1}^{N} i_j e^{i  \omega_0 T\, j},$$ 
which is independent of $z_0$.  Equation (\ref{GeneralMap}) is, then, given by $\mathcal{Z}_{I_N} =\mathcal{R}_{\theta}( \mathcal{T}_{z_0}(\mathcal{W}_{I_N}))$, with $\theta=-N \omega_0 T.$

Finally, we mention that, at least in principle, we may map the results in sections V and VI  to those of a standard Floquet system. If one considers the von Neumann measurement scheme, the microscopic degree of freedom to be measured must become entangled with the apparatus, or pointer states. This can be done via a {\it Hamiltonian} interaction which would lead to the so-called pre-measurement state. In the scenario envisioned in this work, this would have to happen periodically, just like in usual Floquet systems. It turns out that the pre-measurement state suffices to generate the ensamble state (\ref{ReducedStateProtocol}). Of course, it may be a prohibitive task to determine which interaction Hamiltonian should be switched on and off periodically to produce the desired state. 
Note that for results referring to individual realizations, as is the case of the dynamical instability described in Section IV, the collapse from the pre-measurement state to a specific eigenket of the observable is an essential ingredient, which {\it cannot} originate from a Hamiltonian dynamics.
\section{Closing Remarks} 
The fact that the system we introduced is native from the quantum realm and, at the same time, presents evidence of quantum chaos, evades the old notion of quantum chaos as the study of systems which are classically chaotic. Although other systems are known to have this property, in the present case, we report on degrees of freedom that do exist classically, but whose dynamics cannot be paralleled in classical systems. This is not equivalent to observing dynamical sensitivity in degrees of freedom that do not have a classical counterpart, as for instance, spin systems. Conversely, for non-classical degrees of freedom we may have systems that are acknowledgedly non-integrable, presenting, however, a slower than exponential decay in the Loschmidt echo \cite{fine}.

Since coherent states are the closest quantum structures to a point in a classical phase-space, they are a natural and experimentally realizable choice. However, the framework of including ancillary measurements as an active player in the quantum dynamics of a main system is, of course, quite general. It may be the case that, performing invasive measurements on classically non-chaotic systems,  could ``push'' them into a quantum chaotic regime (even after excluding the randomness imparted by the measurements), thus, leading to an alternative route to quantum dynamical instability. Further investigation is needed to verify whether or not this is the case. Note also that the employed procedure is not always equivalent to carry out a series of POVMs in the main system, even if it is time dependent (there is no requirement to resolve the unit operator with the set of possible outcomes). 

In spite of the unstable nature of individual realizations of the system's dynamics, when averages over all possible quantum trajectories are considered, regular patterns emerge. The Husimi function associated with the evolution of arbitrary coherent states may give rise to phase-space crystals, depending on the frequency ratio $R$ and spin-orbit interaction strength. This is an interesting feature because one may reproduce some properties which are typical from a crystalline solid, with a physical system possessing a single potential well.  Although this is an observed trait in archetypal Floquet systems, the mechanism that leads to such structures in the present case is distinct. 

There are several open questions which would be interesting to address. In addition to phase-space crystals, is there a regime where one can observe time crystals \cite{TC} as in other Floquet systems \cite{TC1,TC2,TC3}? Can we observe quantum phase transitions \cite{QPT1,QPT2,QPT3,QPT4} in this class of systems? Finally, we note that fast spin measurements are an important part of the quantum computation program with trapped ions, having many experimental realizations \cite{RMP}. The expertise accumulated in the manipulation of this class of systems makes it a potential candidate for the implementation of the dynamics presented in this work.   
\begin{acknowledgments}
The authors thank Marcelo F. Santos and Eduardo O. Dias for their comments and suggestions on this work. This work received financial support from the Brazilian agencies Coordena\c{c}\~ao de Aperfei\c{c}oamento de Pessoal de N\'{\i}vel Superior (CAPES), Funda\c{c}\~ao de Amparo \`a Ci\^encia e Tecnologia do Estado de Pernambuco (FACEPE), and Conselho Nacional de Desenvolvimento Cient\'{\i}fico  e Tecnol\'ogico through its program CNPq INCT-IQ (Grant 465469/2014-0).
\end{acknowledgments}
\appendix
\section{General expression for the maps $\mathcal{Z}_{I_N}$}
\label{Zmap}
Let $I_N = \{i_N,i_{N-1},...,i_1\}$ represent an arbitrary arrangement of $N$ elements $i_k = \pm 1$. Defining the two complex maps $\mathcal{Z}_{\pm}(z_0) = (z_0 \mp v) e^{-i \omega_0 T} \pm v$, the general expression given by Eq. (\ref{GeneralMap}) holds.

\noindent \textbf{Proof.} Consider the following expressions corresponding to one, two, and three applications of the maps $\mathcal{Z}_{\pm}$, respectively.

\textit{\textbf{n = 1}}
\begin{equation*}
    \begin{split}
        \mathcal{Z}_{i_1}(z_0) &= z_0 e^{-i \omega_0 T} + i v \Big( 1 - e^{-i \omega_0 T} \Big)\\
        &= \left( z_0 + 2i v \sin \qty( \frac{\omega_0 T}{2} ) e^{- i \omega_0 T/2} i_1 e^{i \omega_0 T} \right) e^{- i \omega_0 T}.
    \end{split}
\end{equation*}
\textit{\textbf{n = 2}}
\begin{equation*}
    \begin{split}
        \mathcal{Z}_{i_2} &\qty( \mathcal{Z}_{i_1} ) \equiv \mathcal{Z}_{i_2,i_1}\\
        &= \left( \mathcal{Z}_{i_1} + 2i v \sin \qty( \frac{\omega_0 T}{2} ) e^{- i \omega_0 T/2} i_2 e^{i \omega_0 T} \right) e^{- i \omega_0 T}\\
        &= \Bigg[ \left( z_0 + 2i v \sin \qty( \frac{\omega_0 T}{2} ) e^{- i \omega_0 T/2} i_1 e^{i \omega_0 T} \right) e^{- i \omega_0 T}\\
        &+ 2i v \sin \qty( \frac{\omega_0 T}{2} ) e^{- i \omega_0 T/2} i_2 e^{i \omega_0 T}  \Bigg] e^{- i \omega_0 T}\\
        &= \left( z_0 + 2i v \sin \qty( \frac{\omega_0 T}{2} ) e^{- i \omega_0 T/2} \sum_{j=1}^2 i_j e^{i \omega_0 T j} \right) e^{- 2i \omega_0 T}.
    \end{split}
\end{equation*}

\textit{\textbf{n = 3}}
\begin{equation*}
    \begin{split}
        \mathcal{Z}_{i_3} &\qty( \mathcal{Z}_{i_2,i_1} ) \equiv \mathcal{Z}_{i_3,i_2,i_1}\\
        &= \left( \mathcal{Z}_{i_2,i_1} + 2i v \sin \qty( \frac{\omega_0 T}{2} ) e^{- i \omega_0 T/2} i_3 e^{i \omega_0 T} \right) e^{- i \omega_0 T}\\
        &= \Bigg[ \left( z_0 + 2i v \sin \qty( \frac{\omega_0 T}{2} ) e^{- i \omega_0 T/2} \sum_{j=1}^2 i_j e^{i \omega_0 T j} \right) e^{- 2i \omega_0 T}\\
         &+2i v \sin \qty( \frac{\omega_0 T}{2} ) e^{- i \omega_0 T/2} i_3 e^{i \omega_0 T} \Bigg] e^{- i \omega_0 T}\\
        &= \left( z_0 + 2i v \sin \qty( \frac{\omega_0 T}{2} ) e^{- i \omega_0 T/2} \sum_{j=1}^3 i_j e^{i \omega_0 T j} \right) e^{- 3i \omega_0 T}.
    \end{split}
\end{equation*}
It is easy to recognize a reproducible pattern. By employing the principle of finite induction we suppose that the general expression is valid for
\textit{\textbf{n = N - 1}}: $\mathcal{Z}_{i_{N-1},...,i_1} = \left( z_0 + 2i v \sin \qty( \frac{\omega_0 T}{2} ) e^{- i \omega_0 T/2} \sum_{j=1}^{N-1} i_j e^{i \omega_0 T j} \right) e^{- i (N-1) \omega_0 T}.$
It is clear that it holds for 
\textit{\textbf{n = N}}
\begin{widetext}
\begin{equation*}
    \begin{split}
        \mathcal{Z}_{i_N} \qty( \mathcal{Z}_{i_{N-1},...,i_1} ) &\equiv \mathcal{Z}_{i_N,...,i_1}
        = \left( \mathcal{Z}_{i_{N-1},..,i_1} + 2i v \sin \qty( \frac{\omega_0 T}{2} ) e^{- i \omega_0 T/2} i_N e^{i \omega_0 T} \right) e^{- i \omega_0 T}\\
        &= \Bigg[ \left( z_0 + 2i v \sin \qty( \frac{\omega_0 T}{2} ) e^{- i \omega_0 T/2} \sum_{j=1}^{N-1} i_j e^{i \omega_0 T j} \right) e^{- i (N-1) \omega_0 T} +
        2i v \sin \qty( \frac{\omega_0 T}{2} ) e^{- i \omega_0 T/2} i_N e^{i \omega_0 T} \Bigg] e^{- i \omega_0 T}\\
        &= \left( z_0 + 2i v \sin \qty( \frac{\omega_0 T}{2} ) e^{- i \omega_0 T/2} \sum_{j=1}^N i_j e^{i \omega_0 T j} \right) e^{- i N \omega_0 T}.
    \end{split}
\end{equation*}
\end{widetext}
which finishes the proof.
\section{State of the system: single realization}\label{sec:time-evolution}
\subsection{Time-evolution}
Let $\{\ket{\pm}\}$ and $\{\ket{\pm}_x\}$ denote the eigenstate basis of $\sigma_z $ and $\sigma_x$, respectively. Let $\mathcal{H}_{HO}$ be the Hilbert space of the harmonic oscillator. 
Now suppose that at $t = 0$ we prepare the state of our system as $\ket{\psi} = \ket{\phi} \ket{s}_x$, where $\ket{\phi} \in \mathcal{H}_{HO}$ and $s$ can be either $+1$ or $-1$. Suppose also that the Hamiltonian is taken as
\begin{equation}\label{eq:hamiltonian}
    H = H_{HO} \otimes \mathds{1} - \alpha f(X) \otimes \sigma_z
\end{equation}
where $\alpha$ is a constant with units of energy, $f(X)$ is, for now, an arbitrary dimensionless function of the position operator $X$, $H_{HO} \equiv P^2/2m + m \omega_0 X^2 / 2$ and $\mathds{1}$ denotes the spin unity operator. Defining $U_\pm^T = \exp{ -i H_\pm T/\hbar }$ and $H_\pm = P^2/2m + m \omega_0 X^2/2 \mp \alpha f(X)$, we can prove the following result:
\noindent \textbf \textit{At time $t = T$ the state of the system will be}
\begin{equation}
\frac{1}{2} \Big\{ \Big[ \Big( U_+^T + s U_-^T \Big) \ket{\phi} \Big] \ket{+}_x + \Big[ \Big( U_+^T - s U_-^T \Big) \ket{\phi} \Big] \ket{-}_x \Big\}.
\end{equation}
\noindent \textbf{Proof.} Using  Eq.\eqref{eq:hamiltonian} it is straightforward to show that
\begin{equation*}
    H^k \ket{\phi} \ket{\pm}_x = \Big( H_\pm^k \ket{\phi} \Big) \ket{\pm}
\end{equation*}
where $k$ is an arbitrary integer. In this way, we have
\begin{equation*}
    \begin{split}
        e^{-i H T/ \hbar} \ket{\phi} \ket{\pm} &= \Bigg[ \sum_{k=0}^\infty \qty( \frac{-i T}{\hbar} )^k \frac{H_\pm^k}{k!} \Bigg] \ket{\phi} \ket{\pm}\\
        &= \Big( U_\pm^T \ket{\phi} \Big) \ket{\pm}
    \end{split}
\end{equation*}
and, therefore,
\begin{eqnarray*}
        e^{-i H T/ \hbar} \ket{\phi} \ket{\pm}_x 
        = \frac{1}{\sqrt{2}} \Big[ \Big( U_+^T \ket{\phi} \Big) \ket{+} \pm \Big( U_-^T \ket{\phi} \Big) \ket{-} \Big]\\
        = \frac{1}{2} \Big\{ \Big[ \Big( U_+^T \pm U_-^T \Big) \ket{\phi} \Big] \ket{+}_x + \Big[ \Big( U_+^T \mp U_-^T \Big) \ket{\phi} \Big] \ket{-}_x \Big\},
\end{eqnarray*}
which is the desired result.
\subsection{Time-evolution of a coherent state under the Hamiltonian $H_\pm$}
Let $\ket{z_0} \in \mathcal{H}_{HO}$ be an eigenstate of the operator $a$. Define $b = \sqrt{\hbar/ m \omega_0}$.
If $f(X) = X \sqrt{2}/b$, then $U_\pm^T \ket{z_0} = \ket{\mathcal{Z}_\pm (z_0)}$.

\noindent \textbf{Proof.} Completing the square and defining $X_0 = \alpha \sqrt{2} b/\hbar \omega_0^2$, $H_\pm$ can be rewritten as
\begin{equation*}
    H_\pm = \frac{P^2}{2m} + \frac{m \omega_0^2}{2} \qty( X \mp X_0 )^2 - \frac{m \omega_0^2}{2} X_0^2.
\end{equation*}
That is, up to an additive constant, this is the Hamiltonian of an harmonic oscillator centered at $\pm X_0$. Thus, defining $a_\pm = a \mp v$, where $v = X_0/b\sqrt{2} = \alpha/\hbar \omega_0$.

Let $\{\ket{n}\}$ and $\{\ket{n_\pm}\}$ be the eigenstate basis of $H_{HO}$ and $H_\pm$, respectively. It is well known that $\ket{z_0}$ can be expanded as
\begin{equation*}
    \ket{z_0} = e^{-|z_0|^2/2} \sum_{n=0}^\infty \frac{z_0^n}{\sqrt{n!}} \ket{n}.
\end{equation*}
Now, note the following facts: (i) if $\ket{z_0}$ is an eigenstate of $a$ with associated eigenvalue $z_0$, so it is an eigenstate of $a_\pm \equiv a \mp v$ with associated eigenvalue $z_0 \mp v$; (ii) if $\ket{z_\pm}$ is an eigenstate of $a_\pm$ with associated eigenvalue $z_\pm$, so it is an eigenstate of $a = a_\pm \pm v$ with associated eigenvalue $z_\pm \pm v$. Thus, using fact (i), we may expand
\begin{equation*}
    \ket{z_0} = e^{-|z_0 \mp v|^2/2} \sum_{n_\pm=0}^\infty \frac{(z_0 \mp v)^n_{\pm}}{\sqrt{n_\pm!}} \ket{n_\pm}.
\end{equation*}
Because $\{\ket{n_\pm}\}$ is the eigenstate basis of $H_\pm$ we have that
\begin{equation*}
    U_\pm^T \ket{z_0} = e^{-|z_0 \mp v|^2/2} \sum_{n_\pm=0}^\infty \frac{\qty[(z_0 \mp v) e^{-i \omega_0 T} ]^n_\pm}{\sqrt{n_\pm!}} \ket{n_\pm}.
\end{equation*}
where we have dropped the unimportant (for our purposes) global phase factor $\exp{-i\omega_0 T/2}$. Clearly this is an eigenstate of $a_\pm$ with associated eigenvalue $(z_0 \mp v) \exp{-i\omega_0 T}$. Therefore,
\begin{equation*}
    \begin{split}
        U_\pm^T \ket{z_0} &= \ket{(z_0 \mp v)e^{-i\omega_0 T} \pm v}
        = \ket{\mathcal{Z}_\pm (z_0)},
    \end{split}
\end{equation*}
which ends the proof.
\subsection{Realization of a specific N-step spin measurement trajectory}
Let $\ket{z_i}$, $i = 1,...,n$, be eigenstates of $a$ with associated eigenvalues $z_i$, respectively. If at $t=0$ we prepare the state of our system as $\ket{\psi} = (\sum_{i=1}^n \ket{z_i}) \ket{s}_x$, where, again $s = \pm 1$, the previous results imply that at $t = T$, the state of the system will be given by
\begin{equation*}
    \begin{split}
        \ket{\psi(T)} &= \frac{1}{2} \Bigg\{ \Bigg[ \sum_{i=1}^n \Big( \ket{\mathcal{Z}_+(z_i)} + s \ket{\mathcal{Z}_-(z_i)} \Big) \Bigg] \ket{+}_x \\
        &+ \Bigg[ \sum_{i=1}^n \Big( \ket{\mathcal{Z}_+(z_i)} - s \ket{\mathcal{Z}_-(z_i)} \Big) \Bigg] \ket{-}_x \Bigg\}.
    \end{split}
\end{equation*}
Let us call a \textit{step} a process consisting in (i) letting the system evolve $T$ units of time and (ii) performing a measurement of $S_x$ on the system.

It is clear that a trajectory is characterized by a set of $N$ random outcomes of measurements of $S_x$. Thus, representing the $N$-step trajectory whose set of outcomes is $\{s_1^k,s_2^k,...,s_N^k\}$, $s_j = \pm 1$, by $\gamma_N^k$, and omiting from now on the arguments of all the maps $\mathcal{Z}_{I_N}(z_0)$, we can enunciate the following statement.
If at $t = 0$ the state of the system is prepared as $\ket{\psi_0} = \ket{z_0} \ket{s_0}_x$, then, for the trajectory $\gamma_N^k$, we have a probability
\begin{equation*}
    p[\gamma_N^k] = \frac{1}{4^N} \norm\Big{\sum_{I_N} c_{I_N} \ket{Z_{I_N}}}^2,
\end{equation*}
the corresponding orbital state being
\begin{equation*}
    \ket{\psi_N^k} = \frac{\sum_{I_N} c_{I_N} \ket{\mathcal{Z}_{I_N}}}{\norm\Big{\sum_{I_N} c_{I_N} \ket{\mathcal{Z}_{I_N}}}},
\end{equation*}
where
\begin{equation*}
    c_{I_N} = \prod_{j=1}^N (s_j^k s_{j-1}^k)^{\delta_{-1,i_j}} = \pm 1.
\end{equation*}

\noindent \textbf{Proof.} Let us apply finite induction once again.

\noindent \textbf{Step 1.} At $t=0$ the state of the system is $\ket{\psi_0}$. Then it is let to evolve $T$ units of time so that
\begin{eqnarray*}
    \ket{\psi_0(T)} &=& \frac{1}{2} \Big\{ \Big[ \Big( U_+^T + s_0 U_-^T \Big) \ket{z_0} \Big] \ket{+}_x\\
    &+& \Big[ \Big( U_+^T - s_0 U_-^T \Big) \ket{z_0} \Big] \ket{-}_x \Big\}
\end{eqnarray*}
Then, under a measurement of $S_x$, we have a probability
\begin{equation*}
    p(s_2|s_1,s_0) = \frac{1}{4} \norm\Big{\Big( U_+^T + s_1 s_0 U_-^T \Big) \ket{z_0}}^2
\end{equation*}
that the outcome will be $s_1 \hbar/2$, $s_1 = \pm 1$. The spin state of the system would, therefore, be projected onto the subspace spanned by $\ket{s_1}_x$ and its composite state would become $\ket{\psi_1^k} \ket{s_1}_x$, where
\begin{equation*}
    \ket{\psi_1^k} = \frac{\Big( U_+^T + s_1 s_0 U_-^T \Big) \ket{z_0}}{\norm\Big{\Big( U_+^T + s_1 s_0 U_-^T \Big) \ket{z_0}}}=\frac{\ket{\mathcal{Z}_+} + s_1 s_0 \ket{\mathcal{Z}_-}}{\norm\Big{\ket{\mathcal{Z}_+} + s_1 s_0 \ket{\mathcal{Z}_-}}}.
\end{equation*}
Regarding these results so far, it will prove useful to construct the following table:
\begin{center}
\begin{tabular}{ |c|c| } 
 \hline
 $\mathcal{Z}_{I_N}$ & $c_{I_N}^k$ \\
 \hline
 $\mathcal{Z}_{+}$ & $(s_1 s_0)^0$ \\ 
 $\mathcal{Z}_{-}$ & $(s_1 s_0)^1$ \\
 \hline
\end{tabular}
\end{center}

\noindent \textbf{Step 2.} Using the same reasoning as above, if the system is let to evolve $T$ units of time, its state will be now
\begin{eqnarray*}
    \ket{\psi_1(T)} &=& \frac{1}{2} \Big\{ \Big[ \Big( U_+^T + s_0 U_-^T \Big) \ket{\psi_1^k} \Big] \ket{+}_x\\
    &+& \Big[ \Big( U_+^T - s_0 U_-^T \Big) \ket{\psi_1^k} \Big] \ket{-}_x \Big\}.
\end{eqnarray*}
Now, under a new $S_x$ measurement, we have a probability
\begin{equation*}
    p(s_2|s_1,s_0) = \frac{1}{4} \norm\Big{\Big( U_+^T + s_2 s_1 U_-^T \Big) \ket{\psi_1^k}}^2
\end{equation*}
that the outcome will be $s_2 \hbar/2$, $s_2 = \pm 1$. The spin state of the system would then be projected onto the subspace spanned by $\ket{s_2}_x$ and its composite state would become $\ket{\psi_2^k} \ket{s_2}_x$, where
\begin{equation*}
    \begin{split}
        \ket{\psi_2^k} &= \frac{\Big( U_+^T + s_2 s_1 U_-^T \Big) \ket{\psi_1^k}}{\norm\Big{\Big( U_+^T + s_2 s_1 U_-^T \Big) \ket{\psi_1^k}}}\\
        &= \frac{\Big( U_+^T + s_2 s_1 U_-^T \Big) \Big( U_+^T + s_1 s_0 U_-^T \Big) \ket{z_0}}{\norm\Big{ \Big( U_+^T + s_2 s_1 U_-^T \Big)\Big( U_+^T + s_1 s_0 U_-^T \Big) \ket{z_0}}}
    \end{split}
\end{equation*}
or, more explicitly,
\begin{equation*}
   \ket{\psi_2^k} = \frac{\ket{\mathcal{Z}_{+,+}} + s_1 s_0 \ket{\mathcal{Z}_{+,-}} + s_2 s_1 \ket{\mathcal{Z}_{-,+}} + s_2 s_1 s_1 s_0\ket{\mathcal{Z}_{-,-}}}{\norm\Big{\ket{\mathcal{Z}_{+,+}} + s_1 s_0 \ket{\mathcal{Z}_{+,-}} + s_2 s_1 \ket{\mathcal{Z}_{-,+}} + s_2 s_1 s_1 s_0\ket{\mathcal{Z}_{-,-}}}}.
\end{equation*}
Finally, we get the table
\begin{center}
\begin{tabular}{ |c|c| } 
 \hline
 $\mathcal{Z}_{I_N}$ & $c_{I_N}^k$ \\
 \hline
 $\mathcal{Z}_{+,+}$ & $(s_2 s_1)^0 (s_1 s_0)^0$ \\ 
 $\mathcal{Z}_{+,-}$ & $(s_2 s_1)^0 (s_1 s_0)^1$ \\
 $\mathcal{Z}_{-,+}$ & $(s_2 s_1)^1 (s_1 s_0)^0$ \\
 $\mathcal{Z}_{-,-}$ & $(s_2 s_1)^1 (s_1 s_0)^1$ \\
 \hline
\end{tabular}
\end{center}

\noindent \textbf{Step 3.} Repetition of the same procedure leads to the following table:
\begin{center}
\begin{tabular}{ |c|c| } 
 \hline
 $\mathcal{Z}_{I_N}$ & $c_{I_N}^k$ \\
 \hline
 $\mathcal{Z}_{+,+,+}$ & $(s_3 s_2)^0 (s_2 s_1)^0 (s_1 s_0)^0$ \\
 $\mathcal{Z}_{+,+,-}$ & $(s_3 s_2)^0 (s_2 s_1)^0 (s_1 s_0)^1$ \\
 $\mathcal{Z}_{+,-,+}$ & $(s_3 s_2)^0 (s_2 s_1)^1 (s_1 s_0)^0$ \\
 $\mathcal{Z}_{+,-,-}$ & $(s_3 s_2)^0 (s_2 s_1)^1 (s_1 s_0)^1$ \\
 $\mathcal{Z}_{-,+,+}$ & $(s_3 s_2)^1 (s_2 s_1)^0 (s_1 s_0)^0$ \\
 $\mathcal{Z}_{-,+,-}$ & $(s_3 s_2)^1 (s_2 s_1)^0 (s_1 s_0)^1$ \\
 $\mathcal{Z}_{-,-,+}$ & $(s_3 s_2)^1 (s_2 s_1)^1 (s_1 s_0)^0$ \\
 $\mathcal{Z}_{-,-,-}$ & $(s_3 s_2)^1 (s_2 s_1)^1 (s_1 s_0)^1$ \\
 \hline
\end{tabular}
\end{center}

Inspecting such tables we can infer that the state $\mathcal{Z}_{I_N} = \mathcal{Z}_{\{i_N,i_{N-1},...,i_1\}}$ always comes accompanied by the coefficient
\begin{equation*}
    \begin{split}
        c_{I_N}^k &= a_{i_N,i_{N-1},...,i_1}^k\\
        &= (s_N^k s_{N-1}^k)^{\delta_{-1,i_N}} (s_{N-1}^k s_{N-2}^k)^{\delta_{-1,i_{N-1}}}\dots\\
        &\dots(s_2^k s_1^k)^{\delta_{-1,i_2}} (s_1^k s_0^k)^{\delta_{-1,i_1}}= \prod_{j=1}^N (s_j^k s_{j-1}^k)^{\delta_{-1,i_j}}.
    \end{split}
\end{equation*}
Note that $s_j^k = \pm 1 \Rightarrow c_{I_N}^k = \pm 1$ and that all possible combinations $I_N$ of $N$ elements $i_k = \pm 1$ appear in the summation.

Now, supposing that this is valid for the step $N-1$, where $N$ is an arbitrary integer, we are in a position to generalize the result.

\noindent \textbf{Step N - 1.} We have
\begin{eqnarray*}
    \ket{\psi_{N-2}(T)} &=& \frac{1}{2} \Big\{ \Big[ \Big( U_+^T + s_{N-2} U_-^T \Big) \ket{\psi_{N-2}^k} \Big] \ket{+}_x\\
    &+& \Big[ \Big( U_+^T - s_{N-2} U_-^T \Big) \ket{\psi_{N-2}^k} \Big] \ket{-}_x \Big\},
\end{eqnarray*}
\begin{equation*}
    p(s_{N-1}|s_{N-2},...,s_0) = \frac{1}{4} \norm\Big{\Big( U_+^T + s_{N-1} s_{N-2} U_-^T \Big) \ket{\psi_{N-2}^k}}^2,
\end{equation*}
\begin{equation*}
    \ket{\psi_{N-1}^k} = \frac{\Big( U_+^T + s_{N-1} s_{N-2} U_-^T \Big) \ket{\psi_{N-2}^k}}{\norm\Big{\Big( U_+^T + s_{N-1} s_{N-2} U_-^T \Big) \ket{\psi_{N-2}^k}}}.
\end{equation*}

\noindent \textbf{Step N.}
\begin{eqnarray*}
    \ket{\psi_{N-1}(T)} &=& \frac{1}{2} \Big\{ \Big[ \Big( U_+^T + s_{N-1} U_-^T \Big) \ket{\psi_{N-1}^k} \Big] \ket{+}_x\\
    &+& \Big[ \Big( U_+^T - s_{N-1} U_-^T \Big) \ket{\psi_{N-1}^k} \Big] \ket{-}_x \Big\},
\end{eqnarray*}
\begin{equation*}
    p(s_N|s_{N-1},...,s_0) = \frac{1}{4} \norm\Big{\Big( U_+^T + s_{N} s_{N-1} U_-^T \Big) \ket{\psi_{N-1}^k}}^2,
\end{equation*}
\begin{equation*}
    \begin{split}
        \ket{\psi_{N}^k} &= \frac{\Big( U_+^T + s_{N} s_{N-1} U_-^T \Big) \ket{\psi_{N-1}^k}}{\norm\Big{\Big( U_+^T + s_{N} s_{N-1} U_-^T \Big) \ket{\psi_{N-1}^k}}}\\
        &= \frac{\sum_{I_N} c_{I_N}^k \ket{\mathcal{Z}_{I_N}}}{\norm\Big{\sum_{I_N} c_{I_N}^k \ket{\mathcal{Z}_{I_N}}}}.
    \end{split}
\end{equation*}
Finally, because all measurements are independent from each other, this last state will be achieved with probability
\begin{widetext}
\begin{equation*}
    \begin{split}
        p[\gamma_N^k] &= \frac{1}{4} \frac{\norm\Big{\prod_{j=1}^N \Big( U_+^T + s_j s_{j-1} U_-^T \Big) \ket{z_0}}^2}{\norm\Big{\prod_{j=1}^N \Big( U_+^T + s_j s_{j-1} U_-^T \Big) \ket{z_0}}^2} \times \frac{1}{4} \frac{\norm\Big{\prod_{j=1}^{N-1} \Big( U_+^T + s_j s_{j-1} U_-^T \Big) \ket{z_0}}^2}{\norm\Big{\prod_{j=1}^{N-2} \Big( U_+^T + s_j s_{j-1} U_-^T \Big) \ket{z_0}}^2} \times ...\\
        &\times \frac{1}{4} \frac{\norm\Big{\prod_{j=1}^2 \Big( U_+^T + s_j s_{j-1} U_-^T \Big) \ket{z_0}}^2}{\norm\Big{\Big( U_+^T + s_1 s_{0} U_-^T \Big) \ket{z_0}}^2} \times \frac{1}{4} \norm\Big{\Big( U_+^T + s_1 s_0 U_-^T \Big) \ket{z_0}}^2= \frac{1}{4^N} \norm\Big{ \sum_{I_N} c_{I_N}^k \ket{\mathcal{Z}_{I_N}} }^2.
    \end{split}
\end{equation*}
\end{widetext}
\section{Full ensemble of trajectories: expectation values of physical quantities}\label{ensembleAP}
\subsection{Arbitrary observable}
The following result will be useful in what follows:
Let $A$ be an arbitrary observable. Then, the expectation value of $A$ over the ensemble of trajectories $\{\gamma_N^k\}$ is
\begin{equation}
\label{generalmean}
    \expval{A} = \frac{1}{2^N} \sum_{I_N} \expval{A}{\mathcal{Z}_{I_N}}.
\end{equation}
\noindent \textbf{Proof.} Using proposition 4 we can evaluate the expectation value of $A$ over the trajectory $\gamma_N^k$. That is,
\begin{widetext}
\begin{equation*}
        \expval{A}_N^k = \expval{A}{\psi_N^k}
        = \frac{\sum_{I_N, J_N} c_{I_N}^k c_{J_N}^k \matrixelement{\mathcal{Z}_{I_N}}{A}{\mathcal{Z}_{J_N}}}{\norm\Big{\sum_{I_N} c_{I_N}^k \ket{\mathcal{Z}_{I_N}}}^2}
        = \frac{\sum_{I_N} \matrixelement{\mathcal{Z}_{I_N}}{A}{\mathcal{Z}_{I_N}}}{\norm\Big{\sum_{I_N} c_{I_N}^k \ket{\mathcal{Z}_{I_N}}}^2} + \frac{\sum_{I_N \neq J_N} c_{I_N}^k c_{J_N}^k \matrixelement{\mathcal{Z}_{I_N}}{A}{\mathcal{Z}_{J_N}}}{\norm\Big{\sum_{I_N} c_{I_N}^k \ket{\mathcal{Z}_{I_N}}}^2}.
\end{equation*}
\end{widetext}
where we used $(c_N^k)^2 = 1$.

It is possible to verify that, because each $s_j$ can assume only two values, there is a total of $2^N$ possible trajectories. Thus, if we realize a very large number (ensemble) of such, all of them will be realized with a frequency $p[\gamma_N^k]$, so that
\begin{widetext}
\begin{eqnarray*}
        \expval{A}_N \equiv \sum_{k=1}^{2^N} p[\gamma_N^k] \expval{A}_N^k
        = \frac{1}{4^N} \sum_{k=1}^{2^N} \left( \sum_{I_N} \expval{A}{\mathcal{Z}_{I_N}} + \sum_{I_N \neq J_N} c_{I_N}^k c_{J_N}^k \matrixelement{\mathcal{Z}_{I_N}}{A}{\mathcal{Z}_{J_N}} \right)\\
        = \frac{1}{2^N} \sum_{I_N} \expval{A}{\mathcal{Z}_{I_N}} + \frac{1}{4^N} \sum_{k=1}^{2^N} \sum_{I_N \neq J_N} c_{I_N}^k c_{J_N}^k \matrixelement{\mathcal{Z}_{I_N}}{A}{\mathcal{Z}_{J_N}}.
\end{eqnarray*}
\end{widetext}
Note that, because all trajectories are realized, for each possible trajectory in which, say, $s_j = +1$, there will always exist a ``reciprocal" trajectory in which $s_j = -1$. Thus, if in the former trajectory some coefficient $c_{I_N}^k$ depends on $s_j$, and $c^k_{I_N} = \pm1$, in the latter (i.e. the ``reciprocal" trajectory), we must have $c_{I_N}^k = \mp 1$ (We can see that for the specific coefficient $c_{i_N,...,i_1}^k$, for which $i_k = +1$ for all $k$, is always  to $1$. Nevertheless, this is not relevant for the presented argument). This means that if we expand the summation over $k$ in the second term of the last equation, it will simply vanish, leading to (\ref{generalmean}).
\subsection{Energy}
The expectation value of the energy over an ensemble of trajectories $\{\gamma_N^k\}$ is given by Eq. (\ref{Hclosed}).

\noindent \textbf{Proof.} The previous result implies
\begin{equation*}
    \expval{H}_N = \frac{\hbar \omega_0}{2^N} \left( \abs{\mathcal{Z}_{I_N}}^2 + \frac{1}{2} \right).
\end{equation*}
So, we must evaluate $\sum_{I_N} \abs{\mathcal{Z}_{I_N}}^2$.

Let $\{z_i\}$, $i = 1,...,n$, be an arbitrary set of complex numbers. We know that $|\sum_{j=1}^n z_i|^2 = \sum_{j=1}^n|z_i|^2 + 2 \sum_{i < j} \Re\{z_i z_j^*\}$. Thus, by proposition 1, we have
\begin{equation*}
    \begin{split}
        \abs{\mathcal{Z}_{I_N}}^2- \abs{z_0 }^2 &= 4 v^2 \sin^2 \qty(\frac{\omega_0 T}{2}) \abs{ \sum_{j=1}^N i_j e^{i \omega_0 T j} }^2 + T(i_k)\\
        &= 4 v^2 \sin^2 \qty(\frac{\omega_0 T}{2}) \left( \sum_{j=1}^N \overbrace{|i_j|^2}^{=1} \right) + T(i_k)\\
        &= 4 N v^2 \sin^2 \qty(\frac{\omega_0 T}{2}) + T(i_k)
    \end{split}
\end{equation*}
where $T(i_k)$ is a short notation for all terms which depend linearly on any of the $i_k$'s. Now note the following (and last) fact: if we fix all $i_k$'s, except but one, say, $i_j$, when we expand the summation over $I_N$, there will appear one term for which $i_k = +1$ and another for which $i_k = -1$, so that the two terms shall cancel out. Therefore, we have that $\sum_{I_N} T(i_k) = 0$, and, consequently, we get Eq. (\ref{Hclosed}).
\subsection{Position and momentum}

The expectation value of the position and the momentum over an ensemble of trajectories $\{\gamma_N^k\}$ are
\begin{equation*}
    \expval{X}_N = b \sqrt{2} \qty[ \Re\{z_0\} \cos(\omega_0 T) + \Im\{z_0\} \sin(\omega_0 T) ]
\end{equation*}
and
\begin{equation*}
    \expval{P}_N = \frac{\hbar \sqrt{2}}{b} \qty[ - \Re\{z_0\} \sin(\omega_0 T) + \Im\{z_0\} \cos(\omega_0 T) ].
\end{equation*}
respectively.

\noindent \textbf{Proof.} From the expression for general averages, we get
\begin{equation*}
    \begin{split}
        \expval{X}_N &= \frac{b \sqrt{2}}{2^N} \sum_{I_N} \Re \qty{ \mathcal{Z}_{I_N} }\\
        \expval{P}_N &= \frac{\hbar \sqrt{2}}{2^N b} \sum_{I_N} \Im \qty{ \mathcal{Z}_{I_N} }.
    \end{split}
\end{equation*}
Thus, we must evaluate first $\Re \qty{ \mathcal{Z}_{I_N} }$ and $\Re \qty{ \mathcal{Z}_{I_N} }$:
\begin{equation*}
        \begin{split}
            \Re \qty{ \mathcal{Z}_{I_N} } &= \Re \qty{ z_0 e^{- i N \omega_0 T} } + T(i_k)\\
            \Im \qty{ \mathcal{Z}_{I_N} } &= \Im \qty{ z_0 e^{- i N \omega_0 T} } + T(i_k).
        \end{split}
\end{equation*}
Since $T(i_k)=0$, we get the desired results.
\subsection{Variances of position and momentum}
The variances of the position and the momentum over an ensemble of trajectories $\{\gamma_N^k\}$ are, respectively,
Eq. (\ref{eq:x2_mean}) and Eq. (\ref{eq:p2_mean}).

\textbf{Proof.} We start by writing 
\begin{equation}
    \expval{X^2} = \frac{2 b^2}{2^N} \sum_{I_N} \left( \Re^2 \qty{\mathcal{Z}_{I_N}} + \frac{1}{4} \right),
\end{equation}
\begin{equation}
    \expval{P^2} = \frac{2 \hbar^2}{2^N b^2} \sum_{I_N} \left( \Im^2 \qty{\mathcal{Z}_{I_N}} + \frac{1}{4} \right).
\end{equation}
Let us consider $\Re \qty{\mathcal{Z}_{I_N}}$:
\begin{eqnarray}
\nonumber
    \Re{\mathcal{Z}_{I_N}} &=&\Re{z_0 e^{-i\omega_0T}} \\
\nonumber
    &+& 2 v \sin(\frac{\omega_0 T}{2}) \Re{ i e^{-i (N + 1/2) \omega_0 T} \sum_{j=1}^N i_j e^{ij\omega_0 T} }.
\end{eqnarray}
But, $\qty(\sum_i z_i)^2 = \sum_i z_i^2 + 2 \sum_{i < j} \Re{z_i^* z_j}$ and $\Re{ z_1 z_2 } = \Re{z_1}\Re{z_2} - \Im{z_1} \Im{z_2}$, where $z_i$ are arbitrary complex numbers, imply
\begin{widetext}
\begin{equation}
    \begin{split}
        \Re^2 \qty{ i e^{-i (N + 1/2) \omega_0 T} \sum_{j=1}^N i_j e^{ij\omega_0 T} } &= \Bigg[ \sum_{j=1}^N i_j \Big( - \cos( (N + 1/2)\omega_0 T ) \sin(j \omega_0 T) 
        + \sin( (N + 1/2)\omega_0 T ) \cos(j\omega_0 T) \Big) \Bigg]^2\\
        &= \Bigg[ \sum_{j=1}^N i_j \sin( (N - j + 1/2) \omega_0 T ) \Bigg]^2
        = \sum_{j=1}^N \sin[2]( (N - j + 1/2) \omega_0 T ) + T(i_k)
    \end{split}
\end{equation}
\end{widetext}
($T(i_k)$ has the same meaning as in the proof of propositions 6 and 7). However,
\begin{widetext}
\begin{equation}
    \begin{split}
        \Re^2 \qty{\mathcal{Z}_{I_N}} &= \frac{1}{2 b^2}\expval{\hat{X}}_N^2 + 4 v^2 \sin[2] (\frac{\omega_0 T}{2}) \Re^2 \qty{ i e^{-i (N + 1/2) \omega_0 T} \sum_{j=1}^N i_j e^{ij\omega_0 T} } + T(i_k)\\
        &= \frac{1}{2 b^2} \expval{\hat{X}}_N^2 + 4 v^2 \sin[2](\frac{\omega_0 T}{2}) \sum_{j=1}^N \sin[2]( (N - j + 1/2) \omega_0 T ) + T(i_k).
    \end{split}
\end{equation}
\end{widetext}
Now, defining $N - j + 1/2 = l/2$ (note that $l$ is always odd: $l = 1,3,5,...,2N - 1$), and using $\sum_{I_N} T(i_k) = 0$, we have Eq. (\ref{eq:x2_mean})
which is the desired result.
Using $\Im{ z_1 z_2 } = \Re{z_1}\Im{z_2} + \Im{z_1} \Re{z_2}$, $z_i \in \mathbb{C}$, a similar analysis can be made for the momentum variance, leading to Eq. (\ref{eq:p2_mean}).
%

%


\begin{thebibliography}{99}
\bibitem{Von} In this context, we may refer to a ``von Neumann comb'' analogous to the Dirac comb present in the Hamiltonians of delta-kicked systems.
\end{thebibliography}
\end{document}